\documentclass{aa}  

\usepackage{graphicx}
\usepackage{txfonts}
\usepackage{tabularx}
\usepackage{rotating}     
\usepackage{booktabs}     
\usepackage{multirow}     
\begin{document} 

\title{Planets Around Solar Twins/Analogs (PASTA) II: chemical abundances, systematic offsets, and clues to planet formation}

\titlerunning{Abundance of Solar Twins/Analogs with Planets}
\authorrunning{Sun et al.}

\author{Qinghui Sun\inst{1}\thanks{Corresponding author: \email{qinghuisun@sjtu.edu.cn}}
	\and Chenyang Ji\inst{2}
	\and Sharon Xuesong Wang\inst{2}
	\and Zitao Lin\inst{2}
	\and Johanna Teske\inst{3}
	\and Yuan-Sen Ting\inst{4,5,6}
	\and Megan Bedell\inst{7}
	\and Fan Liu\inst{8,9}
}

\institute{
	Tsung-Dao Lee Institute, Shanghai Jiao Tong University, Shanghai 200240, China\\
	\email{qinghuisun@sjtu.edu.cn}
	\and
	Department of Astronomy, Tsinghua University, Beijing 100084, China
	\and
	Earth and Planets Laboratory, Carnegie Institution for Science, 5241 Broad Branch Road, NW, Washington, DC 20015, USA
	\and
	Department of Astronomy, The Ohio State University, 1251 Wescoe Hall Dr., Columbus, OH 43210, USA
	\and
	Center for Cosmology and AstroParticle Physics, The Ohio State University, 191 West Woodruff Avenue, Columbus, OH 43210, USA
	\and
	Max Planck Institute for Astronomy, Königstuhl 17, D-69117 Heidelberg, Germany
	\and
	Center for Computational Astrophysics, Flatiron Institute, 162 5th Avenue, New York, NY 10010, USA
	\and
	School of Physics and Astronomy, Monash University, Melbourne, VIC 3800, Australia
	\and
	ARC Centre of Excellence for Astrophysics in Three Dimensions (ASTRO-3D), Canberra, ACT 2611, Australia
}


 
\abstract
{Previous studies have suggested that the Sun is relatively depleted in refractory elements compared to other solar twins or analogs, potentially as a result of planet formation. However, such conclusions are often limited by inhomogeneous samples and a lack of direct comparison with stars known to host planets.}
{We aim to perform a homogeneous and precise abundance analysis of solar twins and analogs that host planets, to investigate possible chemical signatures associated with planet formation.}
{We obtain high-resolution, high signal-to-noise ratio Magellan/MIKE spectra for 25 solar-like stars, including 22 confirmed or candidate planet hosts and three comparison stars. Stellar parameters and elemental abundances for 23 elements (from C to Eu) are derived through a strict line-by-line differential analysis relative to the Sun.}
{Our sample spans [Fe/H] = -0.23 to +0.18 dex and includes 20 solar analogs, six of which are solar twins. Typical abundance uncertainties range from 0.01–0.05 dex for lighter elements (e.g., Fe, Si, C, O, Na) and up to 0.1 dex for neutron-capture elements. The Sun is consistently depleted in refractory elements relative to all solar analogs and twins, regardless of planet type. Stars hosting small planets tentatively show slightly stronger refractory element depletion than those hosting giant planets, though the difference is not yet statistically significant.}
{We emphasize the need for strictly differential, line-by-line analyses relative to the Sun, as well as careful consideration of systematic differences between instruments, to ensure consistency and the homogeneity required to achieve our goals.}

   \maketitle
%

\section{Introduction} \label{sec:intro}

The solar chemical composition provides important insight into the formation and evolution of stars and planetary systems, including our own. Previous studies have shown that the Sun is slightly depleted ($<$ 0.1 dex) in refractory elements (with condensation temperatures $T_c > 1300$ K) relative to volatile elements ($T_c \leq 1300$ K), when compared to solar twins and analogs \citep[e.g.][]{2003ApJ...591.1220L, 2009ApJ...704L..66M, 2009A&A...508L..17R, Ramirez2014, 2015A&A...579A..52N, 2016A&A...593A..65N, 2018ApJ...865...68B, 2024ApJ...965..176R, 2025ApJ...980..179S}. The relative abundance depletion can be evaluated by taking the difference between the solar abundance and that of solar twins ([X/Fe]$_{\rm solar}$ – [X/Fe]$_{\rm twin}$), and plotting it as a function of $T_c$ for a range of elements spanning a broad $T_c$ distribution. A negative slope in this trend typically indicates a relative depletion of refractory elements. This trend, initially identified by \citet{2009ApJ...704L..66M}, shows a correlation between differential elemental abundances and $T_c$, suggesting that the Sun is relatively deficient in refractory elements compared to solar twins without detected planets. This result has been confirmed by several subsequent studies (e.g., \citealt{2018ApJ...865...68B, 2021ApJ...907..116N, 2024ApJ...965..176R}) that compares to solar twins/analogs without detected planets, which inspired on linking planet formation to the relative depletion of refractory element in the Sun as the Sun formed planets in the past.

Several planet formation theories have been proposed to explain the possible mechanisms behind the Sun’s relative depletion in refractory elements. In particular, rocky planet formation may lock up refractory material, leaving the host star relatively depleted in these elements \citep{2009ApJ...704L..66M}. \citet{2010ApJ...724...92C} further show that the observed solar refractory element depletion could be explained by the removal of about four Earth masses of rocky material. Similarly, \citet{2020MNRAS.493.5079B} suggest that giant planet formation within the snow line could trap over 100 Earth masses of refractory-rich solids, reducing a host star's refractory element abundance by 5–15\%. Despite those ongoing efforts to develop theoretical models, observational evidence remains limited. 

Most observational studies have focused either on solar twins without detected planets (e.g., \citealt{2018ApJ...865...68B}) or on broader samples of solar analogs (e.g., \citealt{2024ApJ...965..176R}) without stringent selection of solar twins based on high-precision stellar parameters and abundances. The moderate-to-low abundance precision, instrument offsets between studies, and abundance analysis pipelines often introduce systematic uncertainties that hinder robust comparisons across studies. In this context, it becomes important to understand the extent to which instrumental and methodological differences affect derived abundances. In this paper we find that using different spectra or analysis pipelines for the same star can lead to abundance differences of 0.1–0.2 dex, which are comparable to or larger than the subtle abundance trends used to infer planet formation signatures. This highlights the importance of using homogeneous data, consistent methods, and the same instrument, or applying corrections of systematic shifts based on stars in common when different instruments are used, in order to evaluate the subtle chemical differences, particularly in comparisons with planet formation.

An ideal observational evidence would compare the Sun to solar twins both with and without planets. However, currently most solar twins with high-precision abundance measurements do not have detected planets. The Sun is observed to be relatively depleted in refractory elements compared to solar twins without planets based on previous observations. If the Sun shows a similar $T_c$ trend slope to that of planet-hosting solar twins, this would support planet formation as the cause of its refractory depletion. Conversely, if the Sun remains more depleted than both groups while planet-hosting and non-hosting twins show similar slopes, then planet formation is unlikely to explain the Sun’s refractory element pattern. Achieving this goal requires high-resolution, high signal-to-noise spectra to identify solar twins with planets, followed by precise elemental abundance measurements to examine $T_c$ trends. A more detailed comparison between solar twins with and without detected planets, further categorized by planet type and system multiplicity, can place tighter constraints on how planet formation influences stellar composition.

The Planets Around Solar Twins/Analogs (PASTA) survey aims to obtain high-resolution, high signal-to-noise (S/N) spectra for over 100 planet-hosting stars using a consistent observational setup and analysis framework. The primary objectives are to identify and confirm solar twins, derive high-precision elemental abundances for a statistically significant sample of solar twins and analogs, and assess whether planet formation leaves detectable chemical signatures in host stars. In PASTA I, we analyzed 17 solar-like stars, including five solar twins, all confirmed or likely planet hosts. In the present study, we observe 25 additional solar-like stars during one observing night with the MIKE spectrograph on the Magellan II telescope. This new sample includes 22 confirmed or high-probability planet hosts and three comparison stars previously analyzed using HARPS spectra (\citealt{2018ApJ...865...68B}), contributing to the survey's broad goal.

PASTA I primarily established the target list, defined the abundance analysis pipeline, and presented an initial discussion of the survey’s broad goals, but was limited by a relative small sample size. In this work, we significantly expand the sample and explicitly evaluate how the choice of instrument and analysis method impacts the derived stellar abundances. By including stars with existing HARPS-based high-precision abundance measurements, we directly compare results from different instruments and quantify the magnitude of systematic offsets between studies. Combined with the PASTA I sample, we currently have a homogeneous set of 42 solar-like planet hosts, providing improved statistical robustness for testing chemical signatures associated with planet formation.

\section{Target Selection and Observation} \label{sec:target}

Target selection was initially based on a cross-match between Gaia-derived solar analogs and publicly available exoplanet catalogs, including the TESS Objects of Interest (TOIs) and the confirmed planet (KP) host list from the NASA Exoplanet Archive\footnote{Downloaded from \url{https://exoplanetarchive.ipac.caltech.edu/} in September 2024.}. The full candidate list is provided in the appendix of  \citet{2025ApJ...980..179S}. For this work, we prioritized bright stars with high planet-hosting probabilities and stellar parameters close to solar, while excluding the targets already analyzed in \citet{2025ApJ...980..179S}.

\begin{table*}
	\caption{Planetary Parameters\label{tab:planets}}
	\centering
	\small
	\setlength{\tabcolsep}{4pt}
	\begin{tabularx}{\textwidth}{l l X X X c l}
		\hline\hline
		Name$^a$ & $V^a$ (mag) & $P_{\rm orb}^a$ (days) & $R_p^a$ ($R_\oplus$) & $M_p \sin i^a$ or $M_p$ ($M_\oplus$) & WG$^a$ & Reference$^b$ \\
		\hline
		HD 10180 & 7.33 & 1.18, 5.76, 16.36, 49.75$\pm$0.02, 122.72$\pm$0.20, 602$\pm$11, 2248$_{-106}^{+102}$, 67.55$_{-0.88}^{+0.68}$, 9.66$_{-0.07}^{+0.02}$ & -- & 1.39$\pm$0.26, 13.17$\pm$0.63, 11.94$\pm$0.75, 25.4$\pm$1.4, 23.6$\pm$1.7, 21.4$\pm$3.0, 65.3$\pm$4.6, 5.1$_{-3.2}^{+3.1}$, 1.9$_{-1.8}^{+1.6}$ & KP & \citet{2011AA...528A.112L,2012AA...543A..52T} \\
		HD 20782 & 7.40 & 595.86$\pm$0.03 & -- & 1.80$\pm$0.23 & KP & \citet{2006MNRAS.369..249J} \\
		HD 31527 & 7.49 & 16.55, 51.21$_{-0.04}^{+0.04}$, 271.67$_{-2.25}^{+2.11}$ & -- & 10.47$_{-0.87}^{+0.89}$, 14.16$_{-1.23}^{+1.28}$, 11.82$_{-1.64}^{+1.70}$ & KP & \citet{2019AA...622A..37U} \\
		HD 42618 & 6.87 & 148.49$_{-0.18}^{+0.20}$ & -- & 15.2$_{-1.80}^{+1.80}$ & KP & \citet{2021ApJS..255....8R} \\
		HD 45184 & 6.38 & 5.89, 13.14 & -- & 12.19$_{-1.03}^{+1.06}$, 8.81$_{-1.02}^{+1.09}$ & KP & \citet{2019AA...622A..37U} \\
		HD 70573 & 8.67 & 851.8$\pm$11.6 & -- & 1900$\pm$130 & KP & \citet{2007ApJ...660L.145S} \\
		HD 75302 & 7.45 & 4356$_{-112}^{+173}$ & -- & 1716$_{-127}^{+159}$ & KP & \citet{2023AA...678A.107P} \\
		HD 88072 & 7.55 & 18539$_{-3717}^{+8926}$ & -- & 2512$_{-320}^{+485}$ & KP & \citet{2022ApJS..262...21F} \\
		HD 98649 & 8.00 & 5384$_{-274}^{+321}$ & -- & 2110$_{-73}^{+143}$ & KP & \citet{2021AJ....162..266L} \\
		\hline
		TOI-248 & 9.03 & 6.0 & 2.32$\pm$0.99 & -- & PC & -- \\
		TOI-426 & 10.12 & 1.32 & 2.16$\pm$2.42 & -- & PPC & -- \\
		TOI-479 & 11.35 & 2.78 & 12.50$\pm$0.53 & 120.1$\pm$8.6 & KP & \citet{2012AA...544A..72L} \\
		TOI-564 & 11.10 & 1.65 & 11.43$_{-3.25}^{+7.96}$ & 465.1$_{-30.5}^{+31.8}$ & KP & \citet{2020AJ....160..229D} \\
		TOI-652 & 7.93 & 1628$_{-21}^{+22}$ & 2.13$\pm$0.11 & 143$_{-16}^{+13}$ & KP & \citet{2020AJ....160...96T} \\
		TOI-867 & 11.23 & 15.40 & 2.29$\pm$0.35 & -- & PC & -- \\
		TOI 1478 & 10.72 & 10.18 & 11.88$_{-0.44}^{+0.45}$ & 0.85$\pm$0.05 & KP & \citet{2021AJ....161..194R} \\
		TOI-2447 & 10.51 & 69.33 & 9.61$_{-0.12}^{+0.13}$ & 125$_{-9}^{+10}$ & KP & \citet{2024MNRAS.533..109G} \\
		TOI-2479 & 9.93 & 36.84 & 3.79$\pm$0.91 & -- & PC & -- \\
		TOI-2518 & 10.95 & 6.90 & 2.19$\pm$0.15 & -- & PC & -- \\
		TOI-2523 & 10.72 & 4.64 & 2.86$\pm$0.21 & -- & PC & -- \\
		TOI-3359 & 10.22 & 15.81 & 2.24$\pm$0.14 & -- & PC & -- \\
		TOI-4320 & 9.17 & 25.13 & 2.70$\pm$0.09 & $<$17.7 & KP & \citet{2024MNRAS.534.3744N} \\
		\hline
		\multicolumn{7}{c}{Comparison stars from \citet{2018ApJ...865...68B}} \\
		\hline
		HIP 25670 & 8.27 & 1.08 & 0.19$\pm$0.01 & -- & FP & -- \\
		HIP 44713 & 7.31 & -- & -- & -- & -- & -- \\
		HIP 54287 & 7.23 & -- & -- & -- & -- & -- \\
		\hline
	\end{tabularx}
	\tablefoot{
		\tablefoottext{a}{The name, $V$ magnitude, orbital period ($P_{\rm orb}$), planet radius ($R_p$), projected planet mass ($M_p\sin i$) or planet mass ($M_p$), and the TESS Follow-up Observing Program (TFOP) Working Group designation. If a TOI planet is unpublished, values are taken from the TOI catalog. WG labels: ``KP'' = known planet; ``PPC'' = promising planet candidate. If multiple planets are present, values are comma-separated.}\\
		\tablefoottext{b}{References are given only for published planets; others are from catalogs.}
	}
\end{table*}

Observations were carried out on 2024 December 29 using the MIKE spectrograph on the Magellan II (Clay) Telescope at Las Campanas Observatory. We used the standard setup, providing high-resolution ($R \sim 65{,}000$), high signal-to-noise ratio (S/N $\sim$ 200 per pixel at 6000 \AA) spectra. The wavelength coverage spans 3350–5000 \AA\ in the blue arm and 4900–9500 \AA\ in the red.

In total, we observed 25 solar-like stars, including 22 that host confirmed exoplanets or high-probability planet candidates, and three comparison stars previously analyzed by \citet{2018ApJ...865...68B} using HARPS. Among the 22 planet-hosting stars, four have published elemental abundances derived from HARPS spectra, bringing the total number of stars with independent literature values to seven. These stars are used to assess potential offsets related to instrumentation or analysis methods in our abundance measurements. Target properties, including stellar IDs, magnitudes, and planetary parameters, are listed in Table~\ref{tab:planets}.

\section{Stellar Parameters} \label{sec:parameter}

\subsection{Stellar Atmosphere Parameters}

We derive stellar atmospheric parameters following the methods described in \citet{2025ApJ...980..179S}. Briefly, we measure equivalent widths (EWs) for elements with atomic number $Z \leq 30$ using the \textit{splot} task in IRAF, using the line list from \citet{2014ApJ...791...14M}. We keep EW measurements in the range of 10–150 m\AA. Stellar parameters are then determined using the q$^2$ code \citep{Ramirez2014} and MARCS model atmospheres \citep{2008AA...486..951G}, with the Sun adopted as the reference point ($T_{\rm eff}$ = 5777 K, $\log g$ = 4.44 dex, [Fe/H] = 0.0 dex). The effective temperature ($T_{\rm eff}$) is derived by enforcing excitation equilibrium of Fe I lines; surface gravity ($\log g$) is constrained by requiring ionization balance between Fe I and Fe II lines; and microturbulence ($V_t$) is determined by minimizing the trend between Fe I abundances and reduced equivalent widths (log EW/$\lambda$). This process is iterated until all parameters converge.

We estimate uncertainties of stellar atmosphere parameters following an approach similar to those described in \citet{2010ApJ...709..447E, 2025ApJ...980..179S}, accounting for parameter covariances and abundance variations. The complete line list and EW measurements for elements with $Z \leq 30$ are provided in Table \ref{tableA1}, and the final stellar atmosphere parameters including uncertainties are listed in Table \ref{tab:parameters}.

\begin{table*}
	\caption{Stellar parameters.\label{tab:parameters}}
	\centering
	\small
	\setlength{\tabcolsep}{4pt}
	\renewcommand{\arraystretch}{1.1}
	\begin{tabular}{llccccccccccccc}
		\toprule
		Name$^a$ & Age$^a$ & $\sigma_{\rm Age}^a$ & Mass$^a$ & $\sigma_{\rm Mass}^a$ & $M_V^a$ & $\sigma_{M_V}^a$ & Radius$^a$ & $\sigma_{\rm Radius}^a$ & $T_{\rm eff}^b$ & $\sigma_{T_{\rm eff}}^b$ & $\log g^b$ & $\sigma_{\log g}^b$ & $V_t^b$ & $\sigma_{V_t}^b$ \\
		& (Gyr)   & (Gyr)                & ($M_\odot$) & ($M_\odot$) & (mag) & (mag) & ($R_\odot$) & ($R_\odot$) & (K) & (K) & (dex) & (dex) & (km\,s$^{-1}$) & (km\,s$^{-1}$) \\
		\midrule
		HD 10180 & 1.803 & 1.054 & 1.092 & 0.010 & 4.594 & 0.078 & 1.042 & 0.037 & 5956 & 15 & 4.47 & 0.05 & 0.92 & 0.04 \\
		HD 20782 & 5.967 & 1.419 & 0.995 & 0.011 & 4.679 & 0.123 & 1.058 & 0.060 & 5821 & 18 & 4.38 & 0.05 & 0.87 & 0.05 \\
		HD 31527 & 5.515 & 1.481 & 0.981 & 0.012 & 4.682 & 0.112 & 1.028 & 0.053 & 5903 & 18 & 4.40 & 0.05 & 0.99 & 0.04 \\
		HD 42618 & 5.497 & 1.855 & 0.962 & 0.014 & 4.904 & 0.104 & 0.977 & 0.045 & 5767 & 18 & 4.44 & 0.05 & 0.75 & 0.05 \\
		HD 45184 & 1.622 & 0.863 & 1.060 & 0.011 & 4.756 & 0.057 & 0.992 & 0.022 & 5895 & 22 & 4.52 & 0.04 & 0.83 & 0.05 \\
		HD 70573 & 2.312 & 1.577 & 1.028 & 0.022 & 4.805 & 0.123 & 0.976 & 0.045 & 5900 & 52 & 4.79 & 0.12 & 1.28 & 0.11 \\
		HD 75302 & 2.626 & 1.686 & 1.035 & 0.024 & 4.854 & 0.129 & 0.988 & 0.048 & 5803 & 55 & 4.64 & 0.10 & 0.77 & 0.15 \\
		HD 88072 & 3.107 & 1.573 & 1.030 & 0.013 & 4.794 & 0.092 & 1.000 & 0.041 & 5826 & 16 & 4.47 & 0.05 & 0.72 & 0.05 \\
		HD 98649 & 3.787 & 1.881 & 0.974 & 0.015 & 5.012 & 0.086 & 0.940 & 0.035 & 5739 & 22 & 4.50 & 0.05 & 0.77 & 0.06 \\
		TOI-248 & 0.604 & 0.364 & 1.105 & 0.009 & 4.684 & 0.030 & 1.009 & 0.008 & 5930 & 18 & 4.57 & 0.03 & 0.98 & 0.04 \\
		TOI-426 & 1.049 & 0.790 & 1.068 & 0.013 & 4.860 & 0.060 & 0.977 & 0.021 & 5813 & 30 & 4.68 & 0.06 & 0.89 & 0.08 \\
		TOI-479 & 10.527 & 1.819 & 0.900 & 0.011 & 5.029 & 0.121 & 1.007 & 0.055 & 5563 & 19 & 4.38 & 0.05 & 0.55 & 0.07 \\
		TOI-564 & 2.514 & 1.288 & 1.052 & 0.013 & 4.867 & 0.077 & 1.001 & 0.035 & 5732 & 19 & 4.49 & 0.05 & 0.66 & 0.08 \\
		TOI-652 & 2.573 & 1.397 & 1.049 & 0.014 & 4.670 & 0.096 & 1.017 & 0.043 & 5934 & 25 & 4.49 & 0.06 & 0.93 & 0.06 \\
		TOI-867 & 6.172 & 1.258 & 1.002 & 0.012 & 4.605 & 0.130 & 1.085 & 0.065 & 5845 & 19 & 4.36 & 0.05 & 0.93 & 0.05 \\
		TOI-1478 & 6.233 & 1.936 & 0.961 & 0.012 & 4.980 & 0.109 & 0.986 & 0.048 & 5658 & 17 & 4.43 & 0.05 & 0.68 & 0.06 \\
		TOI-2447 & 2.840 & 1.677 & 1.102 & 0.048 & 4.451 & 0.258 & 1.115 & 0.123 & 5997 & 97 & 4.72 & 0.20 & 1.09 & 0.23 \\
		TOI-2479 & 1.819 & 1.244 & 1.031 & 0.014 & 4.892 & 0.068 & 0.960 & 0.026 & 5820 & 28 & 4.55 & 0.05 & 0.78 & 0.06 \\
		TOI-2518 & 1.229 & 0.813 & 1.075 & 0.014 & 4.714 & 0.060 & 0.998 & 0.020 & 5927 & 28 & 4.58 & 0.05 & 0.83 & 0.06 \\
		TOI-2523 & 5.439 & 2.388 & 0.998 & 0.023 & 4.772 & 0.210 & 1.043 & 0.100 & 5763 & 46 & 4.44 & 0.12 & 1.05 & 0.12 \\
		TOI-3359 & 3.103 & 1.436 & 1.074 & 0.020 & 4.609 & 0.133 & 1.068 & 0.064 & 5877 & 27 & 4.44 & 0.07 & 0.91 & 0.07 \\
		TOI-4320 & 0.711 & 0.487 & 1.095 & 0.012 & 4.671 & 0.042 & 1.003 & 0.014 & 5967 & 24 & 4.60 & 0.04 & 0.87 & 0.06 \\
		HIP 25670 & 1.999 & 1.175 & 1.067 & 0.013 & 4.785 & 0.080 & 1.007 & 0.034 & 5815 & 28 & 4.50 & 0.05 & 0.59 & 0.09 \\
		HIP 44713 & 3.216 & 1.178 & 1.092 & 0.014 & 4.571 & 0.131 & 1.099 & 0.069 & 5847 & 20 & 4.40 & 0.06 & 0.83 & 0.05 \\
		HIP 54287 & 2.857 & 1.423 & 1.053 & 0.013 & 4.801 & 0.092 & 1.016 & 0.044 & 5775 & 19 & 4.46 & 0.05 & 0.81 & 0.06 \\
		\bottomrule
	\end{tabular}
	\tablefoot{
		\tablefoottext{a}{Stellar name, age and mass (in solar units) with uncertainties, absolute $V$ magnitude (computed from apparent $V$ and parallax), and stellar radius.}\\
		\tablefoottext{b}{Stellar atmospheric parameters: effective temperature ($T_{\rm eff}$), surface gravity ($\log g$), microturbulence ($V_t$), and their uncertainties.}
	}
\end{table*}

\subsection{Chemical Abundance Analysis}

We perform a strict line-by-line differential abundance analysis relative to the solar spectrum (Vesta, observed with the same Magellan II/MIKE setup), following similar approach described in \citet{2020AJ....159..220S, 2022MNRAS.513.5387S, 2025ApJ...980..179S}. We use EWs with the \textit{abfind} driver in MOOG to derive abundances for each line for each star for elements with atomic number $Z \leq 30$, and employ the \textit{synth} driver for spectral synthesis of heavier elements (Sr, Y, Ba, Eu). For oxygen, we apply NLTE corrections based on the grids from \citet{2007AA...465..271R}. For all other elements, NLTE effects are small due to the strictly line-by-line differential analysis. The line-by-line differential abundance relative to the Sun for each line for each star is provided in Table \ref{tableA2}.

For the same element, we compute the final averaged abundances in the linear space using all lines available for each element after removing $2\sigma$ outliers, and show them in Table\ref{tab:abund_final}. We compute both the standard deviation ($\sigma$) and standard deviation of the mean ($\sigma_{\mu}$ = $\sigma$/$\sqrt{N}$). For element where only one line is used (e.g. K), $\sigma_{\mu}$ directly corresponds to the 1$\sigma$ EW, calculated based on the full width at half maximum, signal-to-noise ratio, and pixel scale of the spectra. We compute the final abundance uncertainty by adding in quadrature the line-to-line scatter ($\sigma_{\mu}$) and the error propagated from uncertainties in atmospheric parameters; the results are shown in Table~\ref{tab:abund_final}.

\begin{table*}
	\caption{Final elemental abundances for each of the 25 stars.\label{tab:abund_final}}
	\centering
	\small
	\setlength{\tabcolsep}{5pt}
	\begin{tabular}{ccccccccc}
		\toprule
		Species$^a$ & Atom$^a$ & 50\% $T_c^a$ & HD~10180$^b$ & Unc.$^b$ & HD~20782$^b$ & Unc.$^b$ & HD~31527$^b$ & ... \\
		&          & (K)          & (dex)        & (dex)    & (dex)         & (dex)    & (dex) & ... \\
		\midrule
		Fe & 26 & 1334 & 0.079  & 0.011 & $-$0.072 & 0.012 & $-$0.223 & ... \\
		C  & 6  &   40 & $-$0.033 & 0.074 & $-$0.229 & 0.035 & $-$0.234 & ... \\
		O  & 8  &  180 & $-$0.001 & 0.024 & $-$0.104 & 0.027 & $-$0.190 & ... \\
		\vdots & \vdots & \vdots & \vdots & \vdots & \vdots & \vdots & \vdots & ... \\
		\bottomrule
	\end{tabular}
	\tablefoot{
		\tablefoottext{a}{Columns 1–3 give the species name, atomic number, and 50\% condensation temperature ($T_c$).}\\
		\tablefoottext{b}{Columns 4–8 list elemental abundances (dex) and associated uncertainties for the 25 stars. The uncertainties include EW measurement errors and those propagated from stellar atmospheric parameter errors.}
		\\ (The full table is available in machine-readable form online.)
	}
\end{table*}

The typical uncertainties (combining contributions from EW measurements and propagated errors from atmospheric parameters) vary across elements. Iron (Fe) and silicon (Si) show the smallest uncertainties, typically 0.011–0.016 dex, though Si reaches 0.024 dex in some stars. For carbon (C), oxygen (O), sodium (Na), magnesium (Mg), and aluminum (Al), uncertainties range from 0.014–0.08 dex, with C showing the widest spread (up to 0.078 dex). Heavier elements like yttrium (Y), barium (Ba), and europium (Eu) have uncertainties of 0.02–0.07 dex, though Ba and Eu occasionally exceed this (0.05–0.11 dex) likely due to fewer detectable lines. Sulfur (S) stands out with the largest uncertainties (0.03–0.28 dex). Sodium (Na) and aluminum (Al) also show significant star-to-star variations (Na: 0.016–0.209 dex; Al: 0.029–0.109 dex), possibly due to line blending or parameter sensitivities.

With final abundance and uncertainty derived, we turn to a detailed comparison between our abundance measurements and previous high-resolution, high S/N studies of the same stars. Rather than repeating the broader survey-level comparisons from PASTA I, we focus here on evaluating potential systematic offsets due to instrument choice and analysis methods, using stars with published HARPS-based abundances.

\subsection{Stellar age, mass, and radius}

We derive stellar ages, masses, and radii using the q$^2$ package \citep{Ramirez2014} with Yonsei-Yale isochrones \citep{2001ApJS..136..417Y}. Our analysis incorporates spectroscopically determined parameters ($T_{\rm eff}$, log g, and [Fe/H]) as primary constraints. This spectroscopic approach provides more reliable results than photometric methods since it is unaffected by interstellar extinction or distance uncertainties \citep{2012AA...543A..29M}. The final derived stellar age, mass, and radius are shown in Table~\ref{tab:parameters}.

\section{Comparison to Literature}

First of all, to verify internal consistency, we measure abundances from multiple solar spectra obtained at different times using Magellan II/MIKE. The resulting abundances show good agreement, confirming the reliability of our analysis method. We also ask different users to measure EWs, which produces slightly different results, mainly due to subjective continuum placement. Some users tend to place the continuum slightly higher or lower, leading to small systematic offsets in EWs and, consequently, in absolute abundances. However, these variations do not affect the relative abundance trends (for example, the slope of abundance versus $T_c$), which are directly used in this study instead of absolute abundances. We emphasizes the importance of maintaining internal consistency and applying a uniform method throughout the analysis.

To assess potential systematic errors, we deliberately include several stars in our sample with existing high-resolution, high S/N abundance measurements for direct comparison. Specifically, we selected three HIP stars without detected planets that were previously analyzed by \citet{2014A&A...572A..48R, 2018ApJ...865...68B} using HARPS spectra (R $\simeq$ 115,000, 3770–6900 \AA\ coverage, S/N $\sim$ 800 pix$^{-1}$ at 6000 \AA). In addition, four planet-hosting stars in our sample (HD 10180, HD 20782, HD 31527, and HD 45184) have published abundance measurements based on HARPS-GTO data \citep{2012A&A...545A..32A, 2017A&A...606A..94D, 2021A&A...655A..99D}, allowing for direct comparison with our MIKE-based results.

Our derived stellar atmosphere parameters agree with previous studies within uncertainties. The mean difference in $T_{\rm eff}$ ($\Delta T_{\rm eff}$) is 42 $\pm$ 23 K (standard deviation, median 47 K), with 5 out of 7 stars within 50 K of the literature values. Surface gravity shows a mean absolute difference of $|\Delta \log g| = 0.07 \pm 0.04$ dex (median 0.06 dex). Metallicity agrees within a mean difference of $0.03 \pm 0.05$ dex (median 0.02 dex), with 6 out of 8 stars within 0.06 dex. Microturbulence shows a slightly larger offset $\Delta V_t = -0.21 \pm 0.09$ km s$^{-1}$, though this has minimal impact on the abundance analysis. 

\begin{singlespace}
	\begin{figure*}
		\centering	\includegraphics[width=0.9\textwidth]{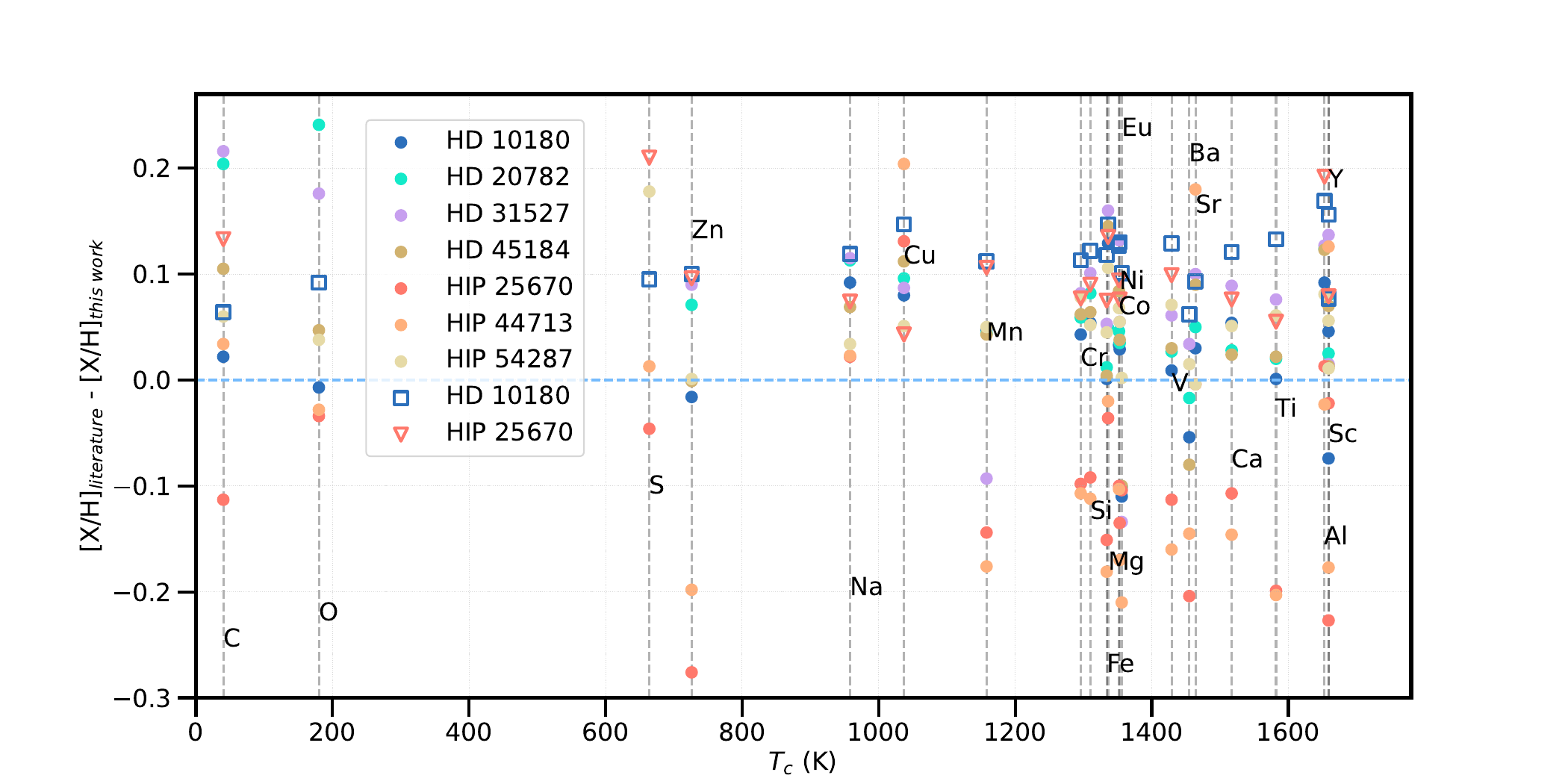}
		\caption{Comparison of elemental abundances from the literature and this work (filled circles) for the same stars. The y-axis shows the difference in abundance, [X/H]${_{\rm literature}}$ – [X/H]${_{\rm this\ work}}$, plotted against the $T_c$ of each element on the x-axis. Potassium (K) is excluded as it was not reported in the literature. The open squares and downward triangles show a direct comparison of abundances measured from HARPS and MIKE spectra using the same spectral lines and analysis methods applied in this study. The differences reflect both instrumental systematics and manual measurement variability between HARPS and MIKE.}
		\label{fig:comp_abund}
	\end{figure*}
\end{singlespace}

Figure~\ref{fig:comp_abund} compares the elemental abundances derived in this work with those reported in the literature, using 1D LTE values for oxygen. Most differences in [X/H] are around 0.1 dex, though some reach up to 0.2–0.3 dex. These moderate systematic offsets are not negligible, especially given our goal of detecting subtle abundance differences between volatile and refractory elements relative to the Sun. Although the abundance offsets tend to be consistent in direction, with values systematically higher or lower across most elements, this consistency may reduce their impact on the derived $T_c$ slope. However, the observed differences still highlight the importance of using homogeneous spectra obtained with the same instrument and analyzed with consistent methods when aiming for high-precision chemical abundance measurements. When combining results from different instruments or studies, it is important to compute and correct for systematic offsets using stars in common. Variations in stellar atmosphere parameters ($T_{\rm eff}$, $\log g$, and $V_t$) typically induce abundance differences of only 0.02–0.04 dex, which are insufficient to explain the larger discrepancies seen between studies.

For the three HIP stars from \citet{2014A&A...572A..48R, 2018ApJ...865...68B}, abundances of elements with atomic number $\le$ 30 are derived from manually measured EWs using the IRAF {\it splot} task, followed by a differential line-by-line analysis similar to ours. For heavier elements (Z $>$ 30), those studies rely on EW measurements, whereas we use synthetic spectrum fitting around each line. This difference in methods likely contributes to the observed discrepancies in [X/H] for those heavier elements. For the four HD stars with known planets (HD 10180, HD 20782, HD 31527, HD 45184), \citet{2012A&A...545A..32A, 2017A&A...606A..94D, 2021A&A...655A..99D} measure EWs using the automated ARES code\footnote{The ARES code can be downloaded at \url{http://www.astro.up.pt/sousasag/ares}.}, and abundances are derived through a similar differential line-by-line analysis. Differences between manual (this study) and automated (ARES) EW measurements may introduce differences in final abundance. Furthermore, all comparison studies use the HARPS spectra, while this work uses Magellan II/MIKE spectra, which differ in wavelength coverage and resolution. These differences lead to the use of different spectral lines and may further contribute to the abundance differences.

Overall, our derived abundances are systematically slightly higher than those reported by \citet{2014A&A...572A..48R, 2018ApJ...865...68B}, but lower than those from \citet{2012A&A...545A..32A, 2017A&A...606A..94D, 2021A&A...655A..99D}. These differences likely arise from a combination of instrumental effects, spectral line selection, analysis methods, and uncertainties in equivalent width (EW) measurements. To quantitatively assess the impact of instrumental shifts and other sources of uncertainty, we compare elemental abundances derived from HARPS and MIKE spectra for two stars, HD 10180 and HIP 25670, using the same linelist and analysis methods.

We measure equivalent widths (EWs) for elements with atomic number $Z < 30$ in the HARPS spectra, which are publicly available from the ESO Archive \citep{2022SPIE12186E..0DR}, using the same line list and analysis procedures as applied to our MIKE spectra, restricted to spectral lines that are present in both instruments. The HARPS spectra do not cover the oxygen (O) triplet at 7771–7776 \AA\ or the potassium (K) line at 7699 \AA, so these lines are excluded from the comparison. Figure~\ref{fig:comp_abund} presents the differential elemental abundances derived from HARPS and MIKE spectra ([X/H]$_{HARPS}$ – [X/H]$_{MIKE}$), using identical spectral lines and analysis methods. Abundances derived from HARPS are consistently higher than those from MIKE. This difference likely arises from a combination of systematic instrumental offsets and variations in manual EW measurements by different users. The remaining discrepancies between our HARPS-based measurements and published literature values likely arise from a combination of differences in the selection of line list and analysis pipelines (e.g., ARES versus manual measurements).

\section{The $T_c$ trend slopes} \label{sec:Tc}

\subsection{Comparing the Sun to solar-like stars}

Our analysis of differential abundance versus $T_c$ trend slopes closely follows the methods of \citet{2025ApJ...980..179S}. Briefly, we calculate [X/Fe] ratios and their differences from solar values ($\Delta$[X/Fe] = [X/Fe]$_{\odot}$ – [X/Fe]$_{\rm star}$), adopting the 50\% equilibrium $T_c$ values from \citet{2003ApJ...591.1220L}. Following standard classifications, we adopt $T_c$ = 1300 K as the boundary between refractory ($T_c$ $>$ 1300 K) and volatile ($T_c$ $<$ 1300 K) elements \citep{2001sse..book.....T, 2021SSRv..217...44L, 2020NatAs...4..314M}.

Figure \ref{figureA1} shows $\Delta$[X/Fe] as a function of $T_c$ for all the 25 solar-like stars. We perform weighted linear regression that accounts for uncertainties (following \citealt{2024AJ....167..167S, 2025ApJ...980..179S}), and display the best-fit slopes with $1\sigma$ confidence intervals in the subplots. The fitted slopes, intercepts, and associated uncertainties are summarized in Table \ref{tab:Tc_slope}. We define solar analogs as stars with $\Delta T_{\rm eff}$ $<$ 200 K, $\Delta \log g$ $<$ 0.2 dex, and $\Delta$[Fe/H] $<$ 0.2 dex relative to the Sun. Twenty stars satisfy these criteria, except for HD 31527, HD 70573, TOI-479, TOI-426, and TOI-2447. We treat the two border lines cases of HD 75302 and HIP 25670 as solar analogs. All stars, except HD 31527 and TOI-479, show negative $T_c$ trend slopes, and thus the Sun is relatively depleted in refractory elements compared to all solar analogs.

The $T_c$ slopes of the 20 solar analogs show good consistency. Notably, the three HIP stars without known planets exhibit $T_c$ slopes consistent with those of the planet-hosting analogs, tentatively supporting a universal depletion pattern that does not depend on the presence of planets. However, small-number statistics remain a limitation. Combined with the 17 stars from PASTA I, we now construct a homogeneous sample of 42 solar-like stars analyzed using consistent observations and techniques.

\begin{table*}
	\caption{$T_{\rm c}$ trend slopes and intercepts.\label{tab:Tc_slope}}
	\centering
	\small
	\setlength{\tabcolsep}{3pt}
	\renewcommand{\arraystretch}{1.2}
	\begin{tabular}{lrrrrrrrrrrrr}
		\hline\hline
		Name &
		\multicolumn{3}{c}{Slope$^a$} &
		\multicolumn{3}{c}{Intercept$^a$} &
		\multicolumn{3}{c}{GCE-corrected Slope$^a$} &
		\multicolumn{3}{c}{GCE-corrected Intercept$^a$} \\
		& Value & $-$err & $+$err & Value & $-$err & $+$err & Value & $-$err & $+$err & Value & $-$err & $+$err \\
		\hline
		HD~10180 & $-$5.5e$-$05 & 2.3e$-$05 & 2.2e$-$05 & 0.089 & 0.029 & 0.030 & -- & -- & -- & -- & -- & -- \\
		HD~20782 & $-$4.4e$-$05 & 3.5e$-$05 & 3.4e$-$05 & 0.106 & 0.044 & 0.045 & $-$4.9e$-$05 & 2.6e$-$05 & 2.6e$-$05 & 0.120 & 0.034 & 0.034 \\
		HD~31527 & 1.7e$-$05 & 1.0e$-$04 & 1.0e$-$04 & $-$0.014 & 1.0e$-$04 & 1.1e$-$04 & -- & -- & -- & -- & -- & -- \\
		HD~42618 & $-$2.5e$-$05 & 3.1e$-$05 & 2.6e$-$05 & 0.059 & 0.034 & 0.040 & -- & -- & -- & -- & -- & -- \\
		HD~45184 & $-$1.1e$-$04 & 3.4e$-$05 & 3.4e$-$05 & 0.178 & 0.044 & 0.044 & -- & -- & -- & -- & -- & -- \\
		HD~70573 & $-$6.1e$-$05 & 5.4e$-$05 & 4.3e$-$05 & 0.083 & 0.053 & 0.068 & -- & -- & -- & -- & -- & -- \\
		HD~75302 & $-$1.5e$-$04 & 4.5e$-$05 & 4.4e$-$05 & 0.237 & 0.057 & 0.057 & -- & -- & -- & -- & -- & -- \\
		HD~88072 & $-$1.0e$-$04 & 2.2e$-$05 & 2.0e$-$05 & 0.157 & 0.026 & 0.027 & $-$1.0e$-$04 & 2.5e$-$05 & 2.3e$-$05 & 0.160 & 0.030 & 0.031 \\
		HD~98649 & $-$6.8e$-$05 & 3.3e$-$05 & 3.3e$-$05 & 0.111 & 0.043 & 0.043 & $-$6.7e$-$05 & 2.8e$-$05 & 2.9e$-$05 & 0.113 & 0.038 & 0.037 \\
		HIP~25670 & $-$6.1e$-$05 & 4.1e$-$05 & 4.0e$-$05 & 0.129 & 0.051 & 0.053 & -- & -- & -- & -- & -- & -- \\
		HIP~44713 & $-$5.3e$-$05 & 4.8e$-$05 & 4.7e$-$05 & 0.133 & 0.059 & 0.062 & -- & -- & -- & -- & -- & -- \\
		HIP~54287 & $-$9.4e$-$05 & 3.3e$-$05 & 3.4e$-$05 & 0.153 & 0.043 & 0.042 & -- & -- & -- & -- & -- & -- \\
		TOI-248   & $-$8.5e$-$05 & 3.1e$-$05 & 3.1e$-$05 & 0.141 & 0.039 & 0.039 & -- & -- & -- & -- & -- & -- \\
		TOI-426   & $-$1.1e$-$04 & 5.6e$-$05 & 5.5e$-$05 & 0.190 & 0.071 & 0.071 & -- & -- & -- & -- & -- & -- \\
		TOI-479   & 2.9e$-$06 & 1.0e$-$04 & 1.0e$-$04 & $-$0.055 & 1.0e$-$04 & 1.1e$-$04 & -- & -- & -- & -- & -- & -- \\
		TOI-564   & $-$9.2e$-$05 & 4.4e$-$05 & 4.4e$-$05 & 0.147 & 0.055 & 0.055 & -- & -- & -- & -- & -- & -- \\
		TOI-652   & $-$4.5e$-$05 & 4.1e$-$05 & 3.8e$-$05 & 0.093 & 0.048 & 0.052 & -- & -- & -- & -- & -- & -- \\
		TOI-867   & $-$2.6e$-$05 & 4.0e$-$05 & 3.6e$-$05 & 0.082 & 0.046 & 0.052 & $-$2.5e$-$05 & 3.4e$-$05 & 3.2e$-$05 & 0.089 & 0.042 & 0.044 \\
		TOI-1478  & $-$5.2e$-$05 & 2.7e$-$05 & 2.7e$-$05 & 0.074 & 0.035 & 0.035 & -- & -- & -- & -- & -- & -- \\
		TOI-2447  & $-$1.1e$-$04 & 3.9e$-$05 & 3.9e$-$05 & 0.166 & 0.050 & 0.050 & -- & -- & -- & -- & -- & -- \\
		TOI-2479  & $-$1.1e$-$04 & 3.5e$-$05 & 3.5e$-$05 & 0.171 & 0.044 & 0.045 & $-$1.1e$-$04 & 3.4e$-$05 & 3.3e$-$05 & 0.166 & 0.043 & 0.044 \\
		TOI-2518  & $-$1.1e$-$04 & 3.2e$-$05 & 3.2e$-$05 & 0.176 & 0.041 & 0.041 & -- & -- & -- & -- & -- & -- \\
		TOI-2523  & $-$3.5e$-$05 & 3.3e$-$05 & 2.8e$-$05 & 0.059 & 0.035 & 0.042 & $-$2.9e$-$05 & 3.0e$-$05 & 2.5e$-$05 & 0.055 & 0.032 & 0.038 \\
		TOI-3359  & $-$1.1e$-$04 & 3.6e$-$05 & 3.6e$-$05 & 0.201 & 0.046 & 0.046 & -- & -- & -- & -- & -- & -- \\
		TOI-4320  & $-$1.2e$-$04 & 4.0e$-$05 & 3.9e$-$05 & 0.195 & 0.050 & 0.051 & -- & -- & -- & -- & -- & -- \\
		\hline
	\end{tabular}
	\tablefoot{
		\tablefoottext{a}{Lower and upper uncertainties for the $T_{\rm c}$ trend slopes and intercepts. For a subset of solar twins, values corrected for Galactic Chemical Evolution (GCE) are also shown.}
	}
\end{table*}

\subsection{Comparing the Sun to solar twins}

We identify six solar twins based on the criteria from \citet{2018ApJ...865...68B}: $\Delta T_{\rm eff} < 100$ K, $\Delta \log g < 0.1$ dex, and $\Delta$[Fe/H] $< 0.1$ dex. We apply Galactic chemical evolution (GCE) corrections to these stars (HD 20782, HD 88072, HD 98649, TOI-867, TOI-2479, and TOI-2523) using the relations from \citet{2018ApJ...865...68B}, except for K, which lacks a correction. Figure~\ref{figureA2} shows the GCE-corrected $T_c$ slopes for the six solar twins. Full regression results are given in Table~\ref{tab:Tc_slope}.

\begin{figure*}
	\centering	\includegraphics[width=0.9\textwidth]{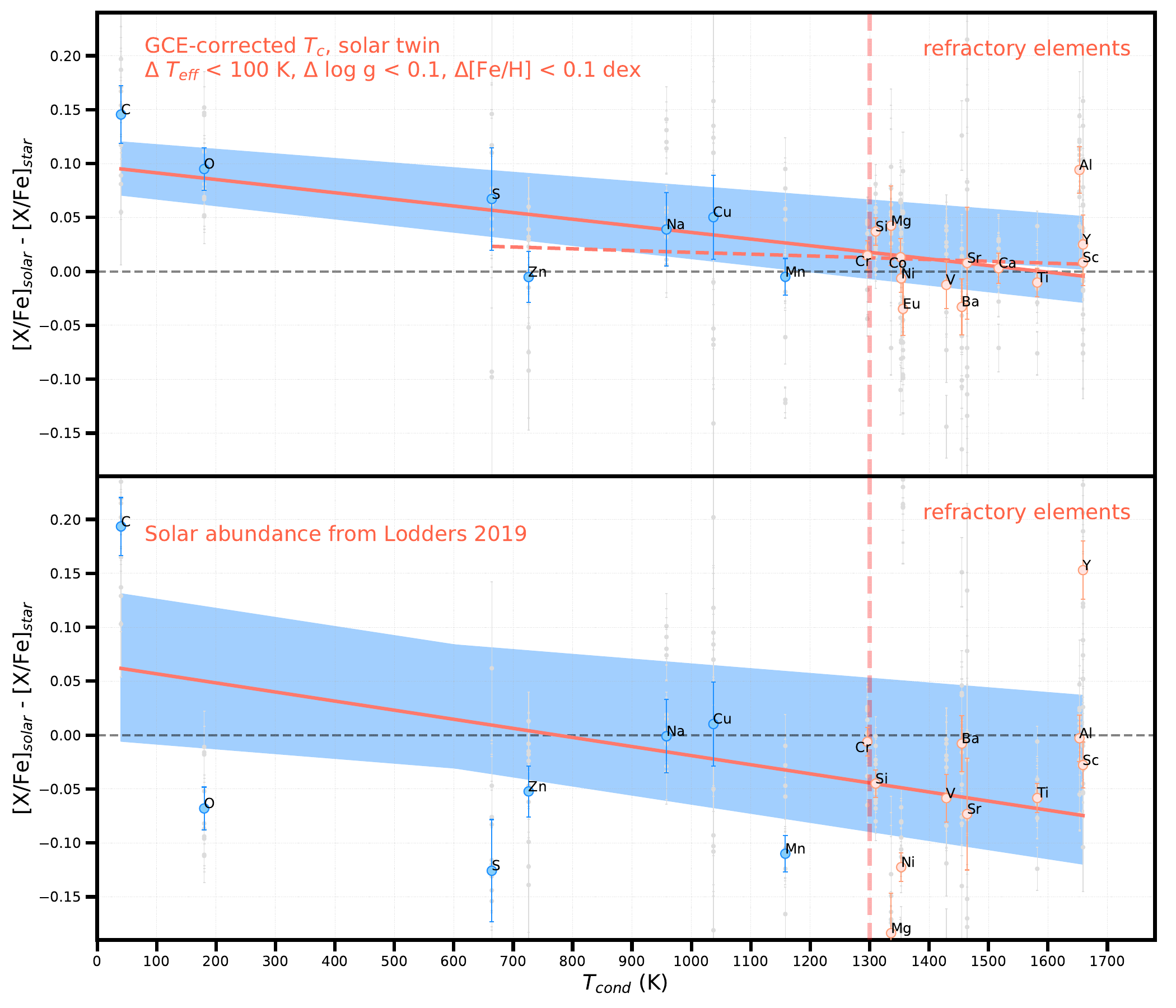}
	\caption{GCE-corrected differential abundances as a function of $T_c$ for 11 solar twins from PASTA I and II. Gray points mark individual stars selected with $-0.1 <$ [Fe/H] $< 0.1$ dex, $\Delta T_{\rm eff} < 100$ K, and $\Delta \log g < 0.1$ dex. Colored symbols represent their mean elemental abundances in linear space: refractory elements ($T_c > 1300$ K) in orange, volatile elements in blue. The vertical dashed orange line indicates the $T_c = 1300$ K boundary. A weighted linear regression to the averaged data (including C and O) is shown as the solid orange line, with a shaded blue band denoting the 1$\sigma$ confidence interval. The regression relation is: $-6.1_{-1.9}^{+1.9} \times 10^{-5} \times T_c + 0.10_{-0.02}^{+0.02}$. A fit excluding C and O is shown as the dashed orange line. The lower panel repeats the analysis using the solar abundances from \citet{2019arXiv191200844L}, yielding: $-8.4_{-5.4}^{+4.0} \times 10^{-5} \times T_c + 0.07_{-0.05}^{+0.07}$.}
	\label{fig:solar twin}
\end{figure*}

Combining these six solar twins with five from the earlier sample, we derive the mean GCE-corrected abundances for 11 planet-hosting solar twins. Their strictly differential abundances relative to the Sun are shown in Figure~\ref{fig:solar twin}, with individual measurements in gray and linear-space averages as colored symbols. Error bars reflect propagated uncertainties. Fe is excluded as it is used as the reference element ([X/Fe]), and K is omitted due to the absence of a GCE correction. We perform MCMC-based linear regression on the mean differential abundances, and found a negative slope is significant at the 3.3$\sigma$ level (${-6.1}_{-1.9}^{+1.9} \times 10^{-05}$), indicating that the Sun is relatively depleted in refractory elements compared to solar twins. Carbon and oxygen contribute substantially to this trend. Excluding C and O results in a slightly shallower, though still negative slope (${-1.7}_{-2.6}^{+1.7} \times 10^{-05}$), further supporting the inferred depletion.

To assess the impact of the differential line-by-line method, we repeat the $T_{\rm c}$ trend analysis using published 1D LTE solar abundances as the reference instead of the strictly differential approach. The lower panel of Figure~\ref{fig:solar twin} shows the same analysis, but adopting the solar abundances from \citet{2019arXiv191200844L}. The corresponding regression yields a steeper slope while remains negative; however, a clear offset is observed between the relation based on literature solar abundances and that derived from our strictly differential analysis. This discrepancy is especially sensitive to C and O abundances, which primarily drive the $T_{\rm c}$ trend in the bottom panel. The comparison highlights the practical importance of performing line-by-line differential abundance analysis using a solar spectrum obtained with the same instrument and setup. By minimizing systematics related to instrumental differences, line selection, and external solar reference scales, this method ensures internally consistent and precise abundance measurements. The comparison with literature solar values confirms the robustness of the negative $T_{\rm c}$ trend, while pointing out the risks of mixing heterogeneous data sources in high-precision studies.

\section{Discussion} \label{sec:discussion}

Our results are consistent with previous findings (e.g., \citealt{2024ApJ...965..176R}) showing that the Sun is relatively depleted in refractory elements compared to solar twins. The origin of this anomaly remains unresolved. We explore two leading hypotheses: the formation of terrestrial planets and the formation of giant planets. In this study, we adopt a practical classification based on planetary radius, where planets with radii less than 7 R$\oplus$ are referred to as ``small planets'' and those with radii greater than or equal to 7 R$\oplus$ as ``giant planets.'' This boundary is not intended to imply a purely rocky or gas-dominated composition, but rather to distinguish between planets likely to differ in their formation and envelope properties. In Table~\ref{tab:planets}, some HD systems provide planetary masses rather than radii; based on these, we classify HD 70573, HD 75302, HD 88072, and HD 98649 as giant-planet hosts, and HD 10180, HD 20782, HD 31527, HD 42618, and HD 45184 as small-planet hosts.

\begin{figure*}
	\centering	\includegraphics[width=0.9\textwidth]{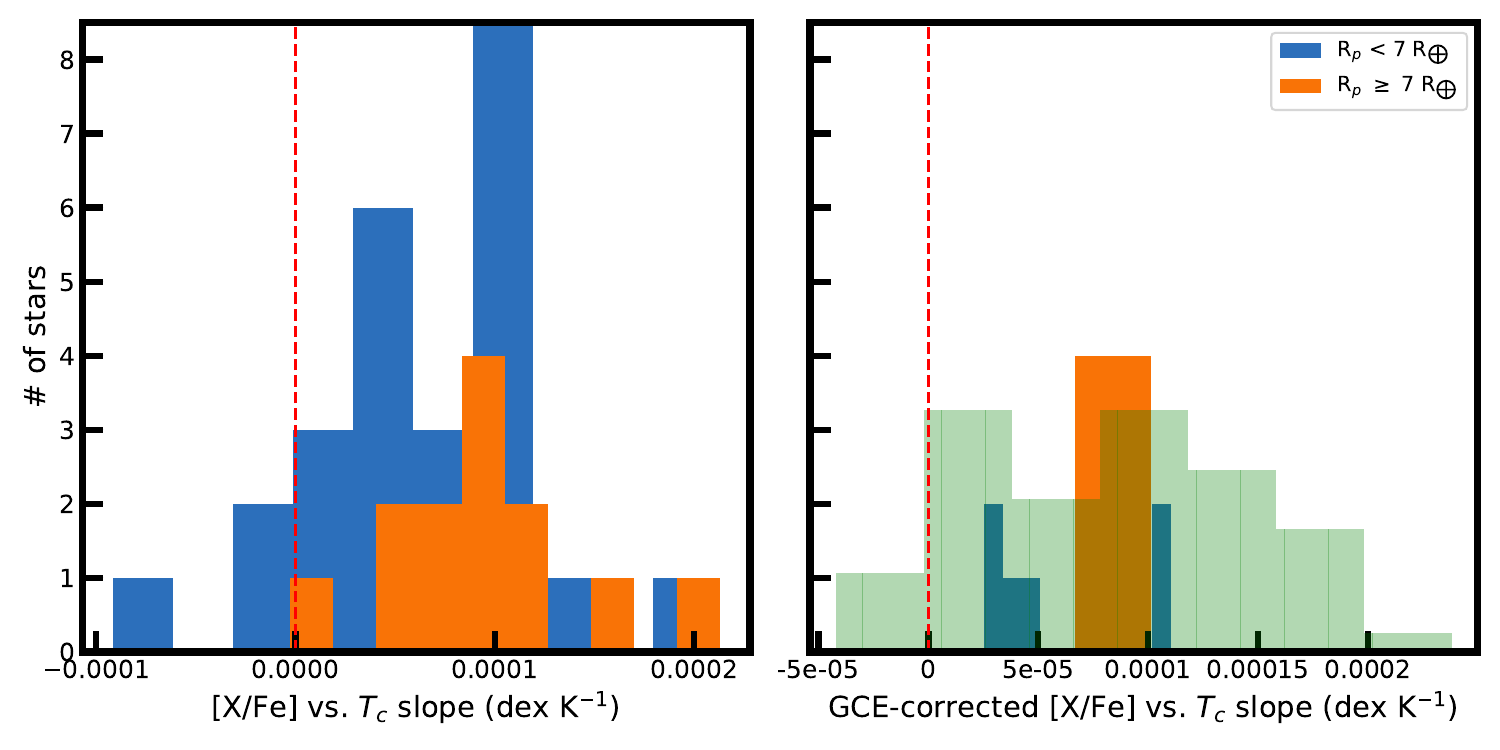}
	\caption{Condensation temperature ($T_c$) trend slopes ([X/Fe]$_{star}$- [X/Fe]$_{\odot}$) for Sun-like stars with different planetary systems. Left: Results for all 39 planet-hosting solar-like stars (this work + PASTA I). Right: Galactic chemical evolution (GCE)-corrected slopes for 11 solar twins. Orange bars indicate stars hosting giant planets ($R_p$ $\ge$ 7$M_{\bigoplus}$), blue bars show systems with only small planets detected ($R_p$ $<$ 7$M_{\bigoplus}$). The Sun's position is marked in a red-dashed line for reference. The background green histogram in the right panel shows the scaled distribution from \citet{2018ApJ...865...68B} for non-planet hosts.}
	\label{fig:Tc_dist}
\end{figure*}

\citet{2009ApJ...704L..66M} proposed that the formation of terrestrial planets can trap refractory elements in the protoplanetary disk, preventing their later accretion onto the host star. Subsequent simulations by \citet{2010ApJ...724...92C} showed that adding approximately four Earth masses of rocky material to the present Sun’s convection zone could reproduce its observed depletion in refractory elements. Regarding giant planet formation, simulations by \citet{2020MNRAS.493.5079B} suggest that early-forming giant planets can trap over 100 Earth masses of dust beyond their orbits, potentially resulting in a 5–15\% depletion of refractory elements in the host star. Nonetheless, observational constraints remain limited due to the small number of solar twins with both precise chemical abundances and confirmed planetary architectures.

To test whether giant or small planet formation can account for the Sun’s refractory element depletion, we compare $T_c$ slopes for systems hosting giant planets ($\geq$7 R$_\oplus$) and those hosting small planets ($<$7 R$_\oplus$)  (Figure~\ref{fig:Tc_dist}). The left panel shows results for all 39 solar-like planet-hosting stars from this work and PASTA I, while the right panel focuses on the 11 GCE-corrected solar twins. In both cases, the $T_c$ slopes of stars with giant planets are similar to those of the broader solar twin population. In contrast, the Sun is more depleted in refractory elements than 36 out of 39 (92.3\%) solar-like stars, including all giant planet hosts, and is also more depleted than all 11 solar twins, regardless of the planet they host. This distribution suggests that giant planet formation is unlikely to explain the Sun’s refractory depletion.

Interestingly, solar twins (Fig. \ref{fig:Tc_dist}, right panel) hosting small planets tentatively show more negative $T_c$ slopes than those with giant planets, suggesting the potential existence of two chemically distinct populations among solar twin planet hosts. This emerging separation has not been identified in earlier studies. To evaluate whether these two populations differ statistically, we perform a Kolmogorov-Smirnov test\footnote{The Kolmogorov-Smirnov (K-S) test compares the cumulative distributions of two samples to assess whether they are drawn from the same parent distribution. A p-value below 0.05 is typically considered statistically significant.} comparing their $T_c$ slope distributions. The test yields a statistic of 0.154 and a p-value of 0.985, indicating no significant difference. Given the limited number of stars in each group, the statistical power remains low, and the current sample size is insufficient to draw firm conclusions. We emphasize the need for additional high-precision abundance analyses of solar analogs and twins with planets in future work. On the other hand, if confirmed, it would suggest that small and giant planet formation affect stellar abundances in different ways, with small planet formation potentially contributing to the Sun’s relative depletion in refractory elements.

\section{Summary} \label{sec:summary}

We obtain high-resolution Magellan II/MIKE spectra for 25 solar-like stars, including 22 with confirmed or high-probability planet candidates and three stars for which their parameters and abundances are previously analyzed in the literature for evaluation of systematic shifts. The sample spans a metallicity range of $-0.23 <$ [Fe/H] $< 0.18$ dex, effective temperatures between $5563 <$ $T_{\rm eff} < 5997$ K, and surface gravities from $4.36 <$ log $g < 4.79$. We then derive precise stellar parameters and elemental abundances for 23 elements, from carbon (C) to europium (Eu). Our strictly differential, line-by-line analysis achieves typical uncertainties below 0.01 dex for well-measured elements such as Fe and Si, and up to 0.04 dex for neutron-capture elements. 

To assess systematic uncertainties between studies, we compare our MIKE-based abundances with previous high-precision results derived from HARPS spectra for stars in common. While our derived stellar parameters agree with literature values within uncertainties, elemental abundances can differ by up to 0.2–0.3 dex in [X/H], primarily due to differences in instrumentation, spectral line selection, and equivalent width measurement techniques. In this work, we explicitly analyze two stars using both HARPS and MIKE spectra, applying the same line list and analysis methods. The HARPS-based abundances are consistently higher than those from MIKE, even under identical procedures, highlighting the importance of using homogeneous datasets in high-precision differential abundance studies. These comparisons underscore the need to correct for instrument- and method-dependent offsets when combining abundance measurements across different surveys.

In this new sample of 25 solar-like stars, we find that the Sun is relatively depleted in refractory elements compared to all solar analogs and solar twins, regardless of whether these stars host giant planets. This conclusion remains valid even when adopting literature solar abundance zero-points; however, the comparison reveals systematic offsets, underscoring the importance of homogeneous, line-by-line differential analysis using the same instrument and setup. We also identify a tentative bimodal distribution in the $T_c$ slopes of stars hosting small versus giant planets, with small planet hosts appearing slightly more depleted in refractory elements. A statistical test, however, does not confirm this difference as significant, but is restricted by the small sample size. This potential separation between the two groups has not been reported in earlier studies. If confirmed, this would suggest that different types of planet formation leave distinct chemical signatures, with small planet formation potentially contributing to the Sun’s refractory depletion.

\section*{acknowledgements}

This work is supported by the National Key R\&D Program of China under Grant Nos. 2025YFE0102100 and 2024YFA1611801. It is also supported by the Science and Technology Commission of Shanghai Municipality under Grant No. 25ZR1402244, and the Shanghai Jiao Tong University New Faculty Startup Program. Y.S.T. is supported by the National Science Foundation under Grant No. AST-240672.

This paper includes data gathered with the 6.5m Magellan Telescopes located at Las Campanas Observatory, Chile, kindly supported by Carnegie Observatories. Some of the targets in this work were observed with the ASTEP telescope. ASTEP benefited from the support of the French and Italian polar agencies IPEV and PNRA in the framework of the Concordia station program and from OCA, INSU, ESA, ERC (grant agreement No. 803193/BEBOP) and STFC; (grant No. ST/S00193X/1).

This paper made use of data collected by the TESS mission and are publicly available from the Mikulski Archive for Space Telescopes (MAST) operated by the Space Telescope Science Institute (STScI). Funding for the TESS mission is provided by NASA’s Science Mission Directorate. We acknowledge the use of public TESS data from pipelines at the TESS Science Office and at the TESS Science Processing Operations Center. Resources supporting this work were provided by the NASA High-End Computing (HEC) Program through the NASA Advanced Supercomputing (NAS) Division at Ames Research Center for the production of the SPOC data products.

\bibliography{sun25_solar}{}
\bibliographystyle{aa}

\appendix

\renewcommand{\thetable}{A\arabic{table}}

\begin{table*}
	\caption{Atomic spectral lines and equivalent widths.\label{tableA1}}
	\centering
	\small
	\setlength{\tabcolsep}{3pt}
	\renewcommand{\arraystretch}{1.1}
	\begin{tabular}{ccccccccccccc}
		\toprule
		$\lambda^a$ [\AA] & Ion$^a$ & $X_{\rm exc}^a$ [eV] & $\log gf^a$ & $C_6^a$ 
		& Solar$^b$ & HD~10180 & HD~20782 & HD~31527 & HD~42618 & HD~45184 & HD~70573 & ... \\
		EW$^c$ [m\AA] & & & & 
		& EW & EW & EW & EW & EW & EW & EW & ... \\
		\midrule
		5044.211 & 26.0 & 2.8512 & $-$2.058 & 2.71E$-$31 & 80.0 & 71.4 & 70.0 & 64.2 & 66.1 & 67.6 & 65.4 & ... \\
		5054.642 & 26.0 & 3.6400 & $-$1.921 & 4.68E$-$32 & 45.3 & 43.6 & 37.2 & 30.7 & 34.2 & 37.0 & 29.9 & ... \\
		\vdots & \vdots & \vdots & \vdots & \vdots & \vdots & \vdots & \vdots & \vdots & \vdots & \vdots & \vdots & ... \\
		\bottomrule
	\end{tabular}
	\tablefoot{
		\tablefoottext{a}{Columns 1–5 give wavelength (in \AA), ion identifier, excitation potential (eV), $\log gf$, and van der Waals damping constant $C_6$. Ion notation: 26.0 = Fe I, 6.0 = C I, etc.}\\
		\tablefoottext{b}{Columns 6–31 give equivalent widths (EW, in m\AA) for the Sun and the 25 target stars, following the header order.}\\
		\tablefoottext{c}{EW: Equivalent width in m\AA.}
		\\ (The full table is available in machine-readable format online.)
	}
\end{table*}

\begin{table*}
	\caption{Abundances derived for each line of each star.\label{tableA2}}
	\centering
	\small
	\setlength{\tabcolsep}{3pt}
	\renewcommand{\arraystretch}{1.1}
	\begin{tabular}{ccccccccccccc}
		\toprule
		$\lambda^a$ [\AA] & Ion$^a$ & $X_{\rm exc}^a$ [eV] & $\log gf^a$
		& HD~10180 & HD~20782 & HD~31527 & HD~42618 & HD~45184 & HD~70573 & HD~75302 & HD~88072 & ... \\
		\midrule
		5044.211 & 26.0 & 2.8512 & $-$2.058 & 0.012 & $-$0.116 & $-$0.202 & $-$0.224 & $-$0.100 & $-$0.309 & 0.034 & $-$0.060 & ... \\
		5054.642 & 26.0 & 3.6400 & $-$1.921 & 0.095 & $-$0.137 & $-$0.249 & $-$0.229 & $-$0.084 & $-$0.302 & 0.111 & $-$0.016 & ... \\
		\vdots   & \vdots & \vdots & \vdots & \vdots & \vdots & \vdots & \vdots & \vdots & \vdots & \vdots & \vdots & ... \\
		\bottomrule
	\end{tabular}
	\tablefoot{
		\tablefoottext{a}{Columns 1–5 list wavelength (in \AA), ion identifier, excitation potential (eV), $\log gf$, and van der Waals damping constant $C_6$. Ion format: atomic number + ionization state (e.g., 26.0 = Fe I).} \\
		\tablefoottext{b}{Columns 6–31 give line-by-line elemental abundances (dex) for the Sun and 25 target stars. Abundances for $Z \leq 30$ are from equivalent widths; for $Z > 30$, spectral synthesis is used.} \\
		(This table is available in full in machine-readable format.)
	}
\end{table*}

\begin{figure*}
	\centering	\includegraphics[width=1.0\textwidth]{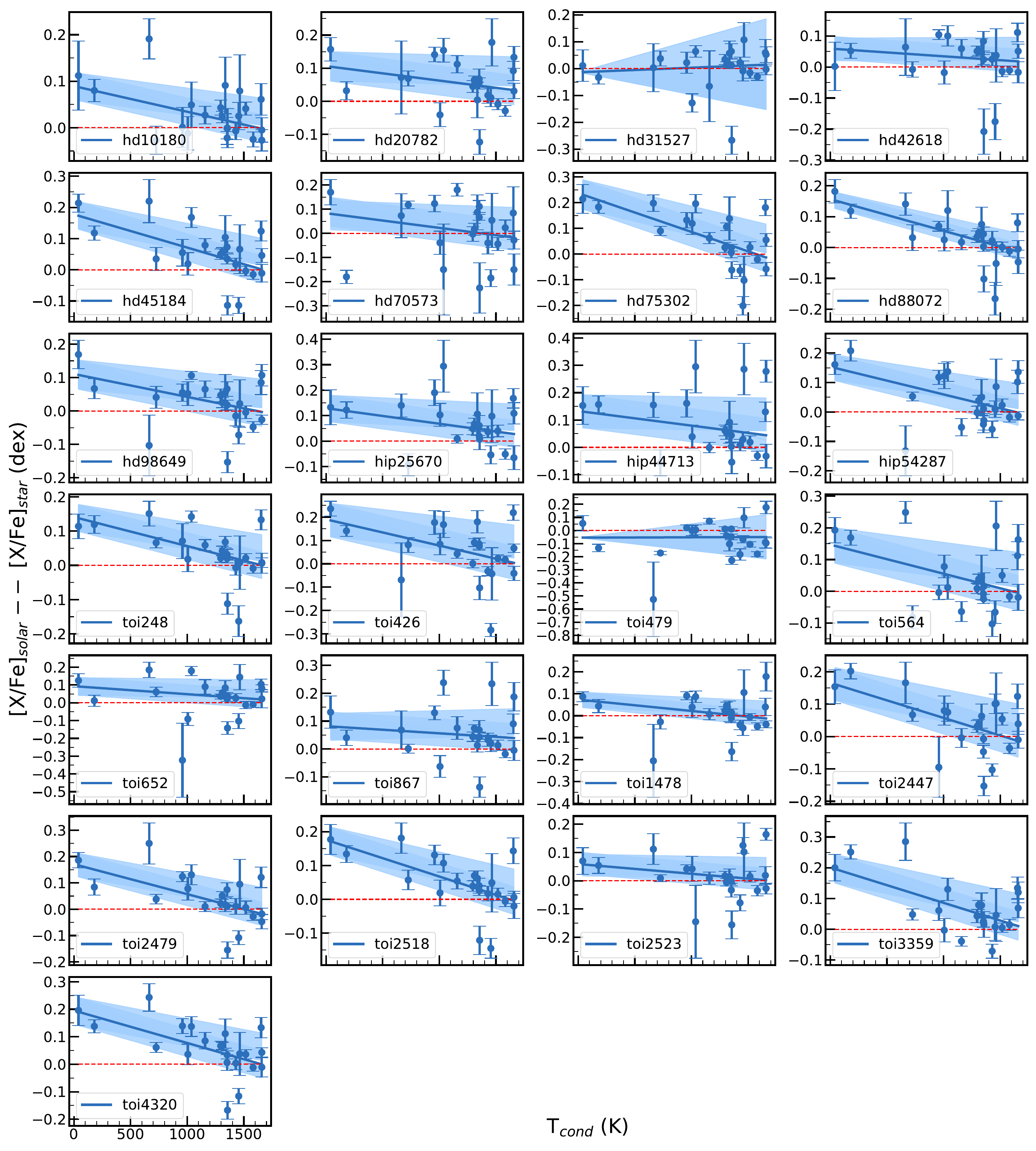}
	\renewcommand\thefigure{A1}
	\caption{The trend of differential elemental abundance ([X/Fe]$_{\rm solar}$ - [X/Fe]$_{\rm star}$) versus $T_c$, with a linear fit to the blue symbols that include error bars, shown by a blue line. The background blue band displays the 1$\sigma$ confidence interval. The Sun is located at a value and slope of 0, shown by the red dashed horizontal line.  \label{figureA1}}
\end{figure*}

\begin{figure*}
	\centering	\includegraphics[width=1.0\textwidth]{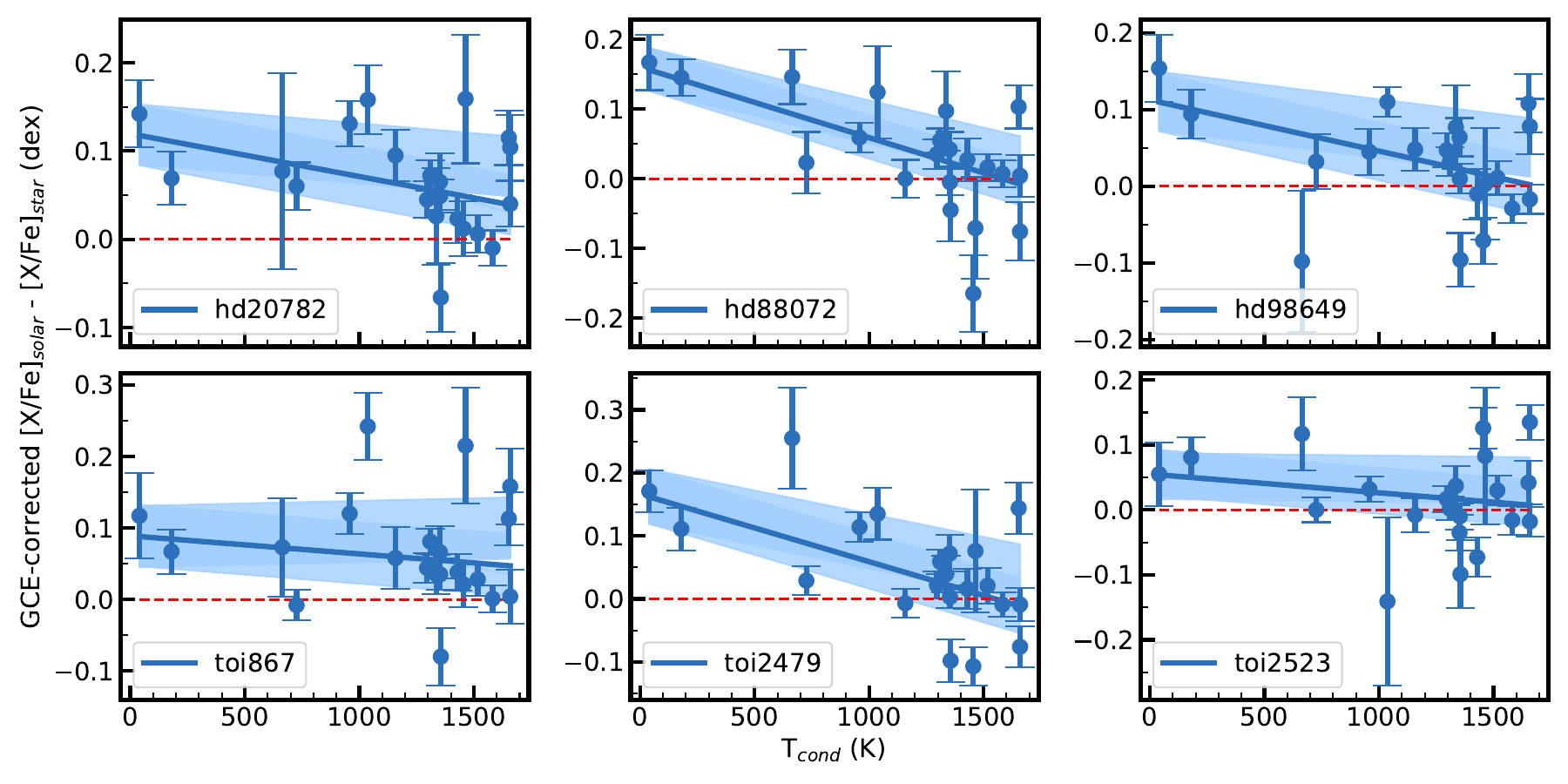}
	\renewcommand\thefigure{A2}
	\caption{The GCE-corrected, differential abundance trend as a function of $T_c$ for the six solar twins. Symbols and interpretations follow those in Figure \ref{figureA1}.  \label{figureA2}}
\end{figure*}

%
%

\end{document}